\documentclass[runningheads]{llncs}
\usepackage{graphicx}
\usepackage{amsmath, amssymb}
\usepackage{xspace, color}

\begin{document}
\title{Facets of Fairness in Search and Recommendation}
%
%
\author{Sahil Verma\inst{1}\and
Ruoyuan Gao\inst{2}\orcidID{0000-0002-8784-4171} \and
Chirag Shah\inst{1}\orcidID{0000-0002-3797-4293}}
%
%
\institute{University of Washington, Seattle WA 98195, USA \\
\email{\{vsahil,chirags\}@uw.edu} \and
Rutgers University, New Brunswick NJ 08901, USA\\
\email{ruoyuan.gao@rutgers.edu}}
\maketitle              
\begin{abstract}
    Several recent works have highlighted how search and recommender systems exhibit bias along different dimensions. 
    Counteracting this bias and bringing a certain amount of fairness in search is crucial to not only creating a more balanced environment that considers relevance and diversity but also providing a more sustainable way forward for both content consumers and content producers. 
    This short paper examines some of the recent works to define relevance, diversity, and related concepts. 
    Then, it focuses on explaining the emerging concept of fairness in various recommendation settings. 
    In doing so, this paper presents comparisons and highlights contracts among various measures, and gaps in our conceptual and evaluative frameworks.
\end{abstract}

\section{Introduction}
Recommendations or ranking candidates for any purpose is an integral part of the technologies we use each day.
Each potential candidate is scored with relevance which is used to rank them in a recommendation list. 
The algorithms used in the underlying software are not only complicated, but they also take clues from the previous actions of the users on the platform. 
This feedback loop potentially leads to discrimination against future users, for example, women less likely to be shown advertisements for high-paying jobs~\cite{womenlessHPJ,guardianwomenlessHPJ}. 

 
Left unchecked, such implicit biases can amount to increased stereotypes and polarized opinions. 
Mitigation of bias in automated decisions is an emerging area in machine learning and related domains. 
Classification and ranking/filtering are the two important categorizations of automated decisions using machine learning. 
Fairness in automated decisions has gained significant traction, and many papers published in the recent years have attempted to 1) devise metrics to quantize fairness, 2) design frameworks using which fair models can be produced (according to the fairness desired metric) or 3) modify data to fight bias in the historical data. 
There have been many works of the kinds mentioned beforehand in both classification and recommendation settings. 
A recent work summarized and explained various fairness metrics used in the classification tasks~\cite{fairness-classification}. 
Unlike classification, recommendations have widely different facets and application scenarios. 
One of the significant differences lies in the output space, which is very restricted in the case of classification. 
In contrast, the output space for ranking or recommendation could be the entire list of ranked items. 
Owing to the sheer abundance of the fairness metrics and their applicability in specific scenarios, understanding their differences and similarities is complicated. 

We review papers from major conferences which received submissions related to fairness in ranking and recommendation, including KDD, WSDM, WWW, SIGIR, ECIR, RecSys, IP\&M, and FAT*, from 2015 to 2020. 
We found twenty-two relevant papers that propose new fairness metrics and provide frameworks to optimize models using them. 
In this paper, we collect and intuitively explain the fairness metrics used in five major recommendation settings: non-personalized recommendation setting, crowd-sourced recommendation setting, personalized recommendation setting, online advertisements and marketplace. 
Since literature has proposed several metrics for each of the settings above, we present all the metrics but we do not attempt to develop arguments in favor of any particular metric, we rather explain the underlying similarities and differences between them, and show how these metrics affect other dimensions of ranking. 
We also develop a clear distinction between fairness in various recommendation settings from often related terms such as diversity, novelty, and relevance. 
We have categorized fairness metrics according to the setting they are applied. 
The main contribution of this work is intuitive categorization and explanation of \textit{twenty-five} fairness definitions in various recommendation settings and identification of relationships between them.

The remainder of this paper is organized as follows. 
Section 2 presents the definitions of commonly used terms in the recommendation literature. 
Sections 3 to 7 delineates the fairness metrics in various settings.
Section 8 outlines the conclusions.

\section{Dimensions of Search and Recommendation results evaluation}
We formally define three dimensions for evaluating search and recommendation results -- relevance, diversity, and novelty. 
These dimensions help to gauge the quality of ranking for a search query or recommendation.     

\begin{itemize}
    \item \textbf{Relevance}~\cite{gao}: Search results are relevant if they accurately answer or describe various aspects of the query or recommendation. 
    It focuses on whether and to what extent a search result relates to the given query. 
    For example, it can be considered as the documents containing some keywords in the query, answering the query, or providing information related to the topic of the query. 
    Relevance only considers the match between the query and the results, often disregarding a user's intent.
    
    \item \textbf{Diversity}~\cite{gao}: Search results for each query might have several topics. For example, given the query ``Lisbon'', the topics include geographic and historical facts, tourism information, and the weather. 
    Diversity refers to the constitution of the search results from its various topics. 
    Various metrics have been defined to measure diversity, such as those found in~\cite{diversity-defs,sakai_simple_2010}.
    
    
    \item \textbf{Novelty}~\cite{gao}: Given several relevant results pertaining to a query, novelty requires the presentation of results that deliver considerably different information content than the results already shown. 
    It thus encourages uniqueness in the shown results with a purpose of maximizing information gain.

\end{itemize}


\section{Fairness Metrics in Non-Personalized Recommendation Settings}
We collect the following metrics from literature to capture fairness in search and recommendations. 
We consider the setting where a ranker wants to rank a set of candidates that are relevant to a query or for recommendation. 
The ranker does not account for individual preferences of the consumers of the ranked list. 
The fairness in this setting addresses how the candidates are ranked. 
We assume the existence of protected and unprotected groups (binary setting), which are defined by law~\cite{protected}.

    \subsection{Accuracy-based fairness metrics}
    Most fairness metrics for recommendations either state the condition for their satisfaction (ideal ranking) or provide a measure of deviation from the ideal ranking. 
    These metrics require a certain proportion of candidates from the protected group in the ranking, or given a ranked list they calculate the divergence from that required proportion in it. 
    Collectively they are called \textit{accuracy-based fairness metrics}. 

    \begin{enumerate}
    \item \textbf{Statistical/demographic Parity: }
    A ranker is said to be satisfying statistical parity if the proportion of candidates from the protected and unprotected groups match the underlying proportion in the top-k rank search results. 
    Therefore statistical parity can be defined at any $k$ length of the ranking. 
    For example, in the image search query, ``CEO'', if the displayed results show an equal proportion of male and female CEOs, the results are said to satisfy statistical parity.
    Singh et al.~\cite{Singh-Joachims} define \textit{exposure} as the resource allocated to a candidate that is computed as a measure of its relevance and position in the ranking. 
    Average exposure to each demographic group also implies statistical parity. 
    If the average exposure to the candidates belonging to different demographic groups is not equal, the ranker is said to violate statistical parity.
    
    \item \textbf{Disparate Treatment: } 
    Disparate treatment refers to the unequal treatment of a candidate due to their membership in a protected group~\cite{disp-treatment}. 
    The candidate would have been treated differently if they had belonged to another group. 
    If the average exposure of different groups is not proportional to their average relevance, then the ranker is said to exhibit disparate treatment~\cite{Singh-Joachims}.
    
    \item \textbf{Disparate Impact: }
    Disparate impact is the practice of not allocating favorable outcomes to protected groups. 
    For searches and recommendations, a click on a candidate is the favorable outcome, which is called as candidate's click-through rate.
    If the expected click-through rate for members of protected and unprotected groups is not proportional to their average relevance, then the ranker is said to cause disparate impact~\cite{Singh-Joachims}.
    Click-through rates for a ranked item can be estimated with the help of several techniques~\cite{click-estimate}.
    
    \item \textbf{Search Neutrality: }
    Search neutrality refers to the search engines' lack of editorial power.
    It means that the search engine should return results to a query impartially and based solely on relevance. 
    It should not promote/demote or differentiate based on websites~\cite{search-neutrality}.
    Google, Facebook, and other tech companies have been accused of violating search neutrality to promote websites that pay them for higher rankings~\cite{google-accused}. 
    
    \item \textbf{Top-k fairness: }~\label{top-k}
    Zehlike et al.~\cite{zehlike2017} describe a ranking as a top-k fair ranking if the top-k candidates in the ranking contain a required proportion of members from the protected group. 
    Given a required proportion, the algorithm they propose ranks candidates in a manner that fairly represents the protected group. 
    Within the protected and unprotected groups, the candidates are ranked by relevance.
    If the required proportion is equal to the underlying proportion between the populations of the protected and unprotected groups, top-k fairness and statistical parity are equivalent.
    
    \item \textbf{Skew@k: }
    Geyik et al.~\cite{Geyik2019} define Skew@k as the logarithm of the ratio of proportions of the candidates belonging to the protected group in the top-k ranked candidates to the desired proportion of the protected group in top-k ranks. 
    A negative skew implies lower than desired representation in the top-k ranks. 
    Zero skew implies a top-k fair ranking (Definition \ref{top-k}).
    Therefore a zero skew with the required proportion equal to underlying proportion would imply equivalence between Skew@k, top-k fair ranking and statistical parity. 
    They also define minskew@k and maxskew@k to extend the metric beyond binary groups. 
    
    \item \textbf{Normalized Discounted Difference (rND)}
    Yang et al.~\cite{julia-paper} point out that establishing statistical parity at higher ranks (e.g., top-10) is more important than establishing it at lower ranks (e.g., top-100). 
    To account for this, they measure set-based fairness metrics at discrete points (e.g., top-10, top-20) and use logarithmic discounts for lower ranks. 
    Normalized discounted difference (rND) computes the difference between the proportion of the protected group members in top-k results and the overall population proportion. 
    If the protected group is proportionally represented, a lower rND value is achieved, which is preferable. 
    If the desired proportion equals to the actual population proportion, then the rND score is correlated with top-k fairness, Skew@k and statistical parity.
    
    \item \textbf{Normalized Discounted KL-divergence (rKL)}
    Normalized discounted KL-divergence(rKL) \cite{julia-paper} measures the expected difference between the proportion of the protected group members in the top-k rank and the overall population proportion. 
    The metric rKL resembles rND, only it is a smoother measure (and therefore optimizable in gradient-based optimization setting) and can be applied to multiple group settings.
    
    \item \textbf{Normalized Discounted Ratio (rRD)}
    Normalized discounted ratio(rRD) \cite{julia-paper} takes the difference between the ratio of the protected to unprotected group members among the top-k ranking and the ratio of underlying sizes of the protected and unprotected groups. 
    A score of zero rRD implies zero skew@k rank, top-k fair ranking and a ranking that satisfies statistical parity. 
    The metric rRD is considered to be useful when the protected group is a minority, in which case it resembles rND and rKL values; otherwise, rRD is meaningless.
    \end{enumerate}

    \subsection{Error based fairness metrics}    
    Kuhlman et al.~\cite{kuhlman_fare_2019} point out that fairness in classification setting has several metrics that are error-based, i.e., they require the classifier to have similar error-rates across the protected and unprotected groups. 
    They claim that those metrics carry value and should be used for measuring fairness in rankings. 
    Unlike a classification task where the error is readily computable, there exists no such error in case of ranking.
    Therefore, Kuhlman et al. propose to use pair-inversions to measure ranking errors.
    They assume the existence of a ground-truth rank for each candidate. 
    If a ranker ranks a candidate higher than its ground-truth rank, they call it a false positive case. 
    Similarly, a candidate that is ranked lower is called a false negative case.
    Each definition that follows has roots in the counterparts described in fairness in classification literature~\cite{fairness-classification}.
    \begin{enumerate}
    \item \textbf{Rank Equality: }
    Rank equality has its origins in the metric called \textit{equalized odds}, which requires equal classification error rates (false positive and false negative error) across the protected and unprotected groups.
    Rank equality error captures the number of times a candidate from a group has been falsely given a higher rank than a candidate of another group; the score is calculated for each such inverted pair.
    This metric does not penalize the ranking where a candidate from the same group has been falsely ranked higher.
    
    \item \textbf{Rank Calibration: }
    Rank calibration~\cite{kuhlman_fare_2019} has roots in calibration which enforces equal precision of classifiers across the protected and unprotected groups.
    It checks how correctly the ranker predicts candidates in each demographic group. 
    Rank calibration error is calculated as the number of times a candidate from one group is falsely ranked higher than candidates of all groups; the score is calculated for each such inverted pair.
    
    \item \textbf{Rank Parity: }
    Rank parity criterion~\cite{kuhlman_fare_2019} has roots in statistical parity. 
    It requires proportional representation of members from the protected and unprotected groups in the ranking.
    The rank parity error is computed as the number of candidates belonging to one group that were ranked higher than candidates from another group; the score is calculated for each such inverted pair.
    \end{enumerate}
    
    \subsection{Causal approach for mitigating discrimination}
    Wu et al.~\cite{Wu-2018} use a causal graph to counteract bias contained in historical data. 
    They use a score variable (instead of rank) to account for individual qualifications and the path-specific effect technique to capture direct and indirect discrimination based on one's membership in the protected group.
    Having detected discrimination, each individual's score is modified to remove the bias, keeping the distribution of new scores close to the original distribution. 
    The modified scores are applied to create a fairer ranking.
    
\section{Fairness Metrics in Crowd-Sourced Non-Personalized Recommendation Settings}
    Chakraborty et al.~\cite{Chakraborty2019} consider the setting of top-k trending recommendations on platforms like Twitter or Yelp which is a non-personalized setting.
    Generally, recommendations are the top-voted candidates using a procedure that resembles an election (with some differences). 
    Each person on the platform can vote (e.g., via a click) for multiple candidates and that too multiple times.
    \begin{enumerate}

    \item \textbf{Equality of Voice: }
    In the setting described above, trends are subject to manipulations by hyper-active group or campaigners of all kinds. 
    This can lead to a veneer of popularity for a particular candidate.
    To avoid this situation, Chakraborty et al. propose a ``one person, one vote'' election procedure in which everyone has an equal say. 
    Each person is asked to specify their preferences across a set of candidates. 
    The first position is assigned to the candidate, which is the first preference of the majority.
    
    \item \textbf{Proportionality for Solid Coalitions: }
    Chakraborty et al.~\cite{Chakraborty2019} point out that due to the abundance of options, user's votes might get split across irrelevant or redundant alternatives, e.g. if there are three candidates out of which the first two are very close. 
    Assume that 60\% of the people are interested in the first two candidates. 
    Due to their similarity, votes would split among them. 
    Thus, even though the sum of the votes across these two candidates is more, a less popular third candidate would emerge as the winner.
    To avert this, proportionality for solid coalitions requires the diversity of opinions in the overall population should be proportionally represented in the top-k recommendations.
    
    \item \textbf{Anti-Plurality: }
    Chakraborty et al. also propose that if a majority of users dislike a candidate, it should not be in the top-k recommendations. 
    In the previous example, the third candidate, disliked by 60\% of the population, would not be recommended.
    \end{enumerate}

\section{Fairness Metrics in Personalized Recommendation Settings}
    Beutel et al.~\cite{beutel_fairness_2019} consider fairness metrics for personalized recommendation settings. 
    Consider $M$ total candidates, out of which $M'$ are relevant to a query, but only $K$ candidates are useful as part of personalization.
    Since they are dealing with personalized ranking along with clicking, Beutel et al. also consider the engagement of the user with a recommended candidate. 
    Engagement between a given user and recommended candidate can be estimated. 
    They compare ranked candidates pairwise and define \textbf{pairwise accuracy} as the probability that a clicked candidate is ranked above another relevant unclicked candidate. 
    \begin{enumerate}
    \item \textbf{Pairwise Fairness: }
    A ranker is said to satisfy pairwise fairness if the probability of a clicked candidate being ranked higher than another relevant unclicked candidate is the same across groups, conditioned on the candidates that have the same predicted engagement score.
    Pairwise fairness does not eliminate systematic preference between demographic groups. 
    For example, one can rank all candidate belonging to a favored group that are not relevant to the query and give a lower rank to candidates from the other group that are relevant to the query. 
    
    \item \textbf{Inter-Group Pairwise Fairness: }
    A ranker satisfies inter-group pairwise fairness~\cite{beutel_fairness_2019} if the probability of a clicked candidate being ranked higher than a relevant but unclicked candidate in the other group is the same across pairs of demographic groups, conditioned on the candidates that have same engagement score.
    
    \item \textbf{Intra-Group Pairwise Fairness: }
    A ranker is said to satisfy intra-group pairwise fairness~\cite{beutel_fairness_2019} if the probability of a clicked candidate being ranked higher than another relevant but unclicked candidate in the same group is equal across demographic groups, conditioned on the candidates that have the same predicted engagement score.
    \end{enumerate}
    \noindent A combination of Intra-Group and Inter-Group Pairwise Fairness can reduce systematic bias against a demographic group.

\section{Fairness Metrics in Advertisement Settings}      
    Chawla et al.~\cite{chawla2019multicategory} present fairness concerns from an entirely different perspective. 
    Thus far, we have described metrics that view fairness from the perspective of a candidate to be ranked or recommended.
    Chawla et al. describe fairness from the perspective of individuals who are being served advertisements.  
    Individual fairness~\cite{fairness-classification} states that the advertisements shown to two similar individuals should be similar.
    For instance, qualifications of an individual can characterize the similarity.
    Nevertheless, the solution to the problem does not lie in showing an equal proportion of advertisements from all categories to similar people, as individuals have needs and preferences. 
    
    \begin{enumerate}

    \item \textbf{Envy-Freeness: }
    Envy-freeness~\cite{chawla2019multicategory,serbos-fairness-2017} is a complementary notion to individual fairness, in which only a user's preference is considered for advertisements and a user's qualifications (therefore, similarities) are not reasoned.
    It requires every user to be content with their share of advertisements. 
    
    \item \textbf{Inter-Category Envy-Freeness: }
    Enforcing individual fairness for all categories of advertisements is problematic since it does not recognize individual preferences. 
    Inter-category envy-freeness~\cite{chawla2019multicategory} allows each user to specify individual preferences, and the criterion requires that all users interested in a category should be served the same amount of advertisements belonging from that category. 
    For example, two individuals interested in jobs should be shown the same number of job-related ads.
    
    \item \textbf{Total Variation Fairness: }
    Inter-category envy-freeness does not ensure a fair distribution of advertisements within a category.
    For instance, two equally qualified individuals belonging to different demographic groups can be unfairly shown high-paying and low-paying jobs respectively, while satisfying that metric.
    Total variation fairness~\cite{chawla2019multicategory} overcomes this limitation by requiring that all subsets of the advertisements from any category shown to two similar individuals must be the same. 
    Consequently, it evades the problem of unfairly showing high-paying job ads to one user.
    
    \item \textbf{Compositional Fairness: }
    Compositional fairness~\cite{chawla2019multicategory} combines inter-category envy-freeness and total variation fairness. 
    Compositional fairness has two requirements: 1) a user's preferences must be recognized, and advertisements served to them belong to their preferred categories only (\textit{envy-freeness}) and 2) within each category, the proportions of advertisements should be the same across all users interested in that category. 
    This lets an advertiser serve advertisements from different categories with varying probabilities to a user (based on their preference). 
    However, the mix of advertisements from each category should be the same across all interested users. 
\end{enumerate}

\section{Fairness Metrics in Marketplace Settings}

    Advertisement setting brings us to a discussion related to marketplaces, in which consumers are shown advertisements about products which the suppliers want to publicize. 
    Marketplaces are ubiquitous. 
    Almost all online platforms we interact with serve as a marketplace for consumers and service providers. 
    These multi-sided recommendation platforms have complicated fairness constraints. 
    Historically, most marketplaces have optimized for consumer satisfaction, but given the rising competition among different platforms, the satisfaction of service providers has also gathered attention. 
    For example, Spotify would like to recommend tracks that a particular consumer would find relevant and is likely to listen. 
    But it would be problematic if Spotify only recommends songs from a few popular artists to the consumers because: 1) it gives low exposure to less popular artists, and 2) the consumer may not find the recommendations interesting after sometime. 
    There has been a few recent works discussing and addressing these concerns~\cite{mehrotra_towards_2018,abdollahpouri_multi-stakeholder_2019,wan_addressing_2020,ferraro_artist_2019,abdollahpouri_recommender_2017,burke_multisided_2017}.
    
    There are several classes of multi-sided recommendation: 1) Multi-receiver recommendation -- when a target audience is a group of people rather than an individual, e.g., students on an education platform; 2) Multi-provider recommendation -- when several suppliers provide the recommendation content, and the platform needs to choose between them, e.g., Airbnb and Spotify, and 3) Side stakeholder recommendation -- when there are parties other than suppliers and consumers involved in the marketplace, the recommendations need to consider their preferences as well, e.g., drivers in the Uber Eats platform.
    
    \begin{enumerate}
        \item \textbf{Consumer Fairness: } 
        A recommendation satisfies consumer fairness~\cite{abdollahpouri_multi-stakeholder_2019} it is does not cause any disparate impact on members of protected groups. 
        For example, consumers of all groups should be served the same distribution of job ads. 
        
        \item \textbf{Provider Fairness: }
        A recommendation satisfies provider fairness~\cite{abdollahpouri_multi-stakeholder_2019} if all the providers have an equal chance of exposure to the consumers. 
        For example, Spotify recommends both famous and less famous artists publishing a specific genre of music to users who prefer that genre.
        
        \item \textbf{Side Stakeholder Fairness: }
        A recommendation satisfies side stakeholder fairness~\cite{abdollahpouri_multi-stakeholder_2019} if it takes into consideration the preferences of side stakeholders. 
        For example, fairly distributing consumer orders and commute distance among drivers in Uber Eats. 
    \end{enumerate}

\section{Conclusion}
In this short survey, we collect and present various metrics proposed in the emerging literature on fairness in recommendations. 
Succinct and distinct categorization of fairness metrics would help people understand the landscape and triage missing gaps, consequently fuelling future research. 
We are already experiencing an adoption of this research in the industry. For instance, 
Geyik et al.\cite{Geyik2019}, in a first large-scale deployment, enforced fair ranking in Linkedln search. 
We envision such deployments to other major search engines in the future.



\bibliographystyle{splncs04}
\bibliography{refs}

\begin{thebibliography}{10}
\providecommand{\url}[1]{\texttt{#1}}
\providecommand{\urlprefix}{URL }
\providecommand{\doi}[1]{https://doi.org/#1}

\bibitem{protected}
\textit{Protected Group}.
  \url{{https://en.wikipedia.org/wiki/Protected\_group}}, accessed: 2020-01-20

\bibitem{google-accused}
\textit{What is Search Neutrality?}
  \url{https://hackernoon.com/what-is-search-neutrality-d05cc30c6b3e},
  accessed: 2020-01-20

\bibitem{guardianwomenlessHPJ}
\textit{Women less likely to be shown ads for high-paid jobs on Google, study
  shows}.
  \url{https://www.theguardian.com/technology/2015/jul/08/women-less-likely-ads-high-paid-jobs-google-study},
  accessed: 2020-01-20

\bibitem{abdollahpouri_recommender_2017}
Abdollahpouri, H., Burke, R., Mobasher, B.: Recommender systems as
  multistakeholder environments. In: Proceedings of the 25th Conference on User
  Modeling, Adaptation and Personalization. UMAP ’17, Association for
  Computing Machinery, New York, NY, USA (2017). \doi{10.1145/3079628.3079657}

\bibitem{abdollahpouri_multi-stakeholder_2019}
Abdollahpouri, H., Burke, R.D.: Multi-stakeholder recommendation and its
  connection to multi-sided fairness. ArXiv  \textbf{abs/1907.13158} (2019)

\bibitem{diversity-defs}
Amig\'{o}, E., Spina, D., Carrillo-de Albornoz, J.: An axiomatic analysis of
  diversity evaluation metrics: Introducing the rank-biased utility metric. In:
  The 41st International ACM SIGIR Conference on Research \& Development in
  Information Retrieval. SIGIR ’18, Association for Computing Machinery, New
  York, NY, USA (2018). \doi{10.1145/3209978.3210024}

\bibitem{beutel_fairness_2019}
Beutel, A., Chen, J., Doshi, T., Qian, H., Wei, L., Wu, Y., Heldt, L., Zhao,
  Z., Hong, L., Chi, E.H., Goodrow, C.: Fairness in recommendation ranking
  through pairwise comparisons. In: Proceedings of the 25th ACM SIGKDD
  International Conference on Knowledge Discovery \& Data Mining. KDD ’19,
  Association for Computing Machinery, New York, NY, USA (2019).
  \doi{10.1145/3292500.3330745}

\bibitem{burke_multisided_2017}
Burke, R.: Multisided {Fairness} for {Recommendation}. arXiv:1707.00093 [cs]
  (Jul 2017), arXiv: 1707.00093

\bibitem{Chakraborty2019}
Chakraborty, A., Patro, G.K., Ganguly, N., Gummadi, K.P., Loiseau, P.: Equality
  of voice: Towards fair representation in crowdsourced top-k recommendations.
  In: Proceedings of the Conference on Fairness, Accountability, and
  Transparency. FAT* ’19, Association for Computing Machinery, New York, NY,
  USA (2019). \doi{10.1145/3287560.3287570}

\bibitem{womenlessHPJ}
Datta, A., Tschantz, M.C., Datta, A.: Automated experiments on ad privacy
  settings: A tale of opacity, choice, and discrimination. ArXiv
  \textbf{abs/1408.6491} (2014)

\bibitem{ferraro_artist_2019}
Ferraro, A., Bogdanov, D., Serra, X., Yoon, J.J.: Artist and style exposure
  bias in collaborative filtering based music recommendations. ArXiv
  \textbf{abs/1911.04827} (2019)

\bibitem{gao}
Gao, R., Shah, C.: Toward creating a fairer ranking in search engine results.
  Information Processing \& Management  \textbf{57} (2020).
  \doi{10.1016/j.ipm.2019.102138}

\bibitem{Geyik2019}
Geyik, S.C., Ambler, S., Kenthapadi, K.: Fairness-aware ranking in search \&
  recommendation systems with application to linkedin talent search. In:
  Proceedings of the 25th ACM SIGKDD International Conference on Knowledge
  Discovery \& Data Mining. KDD ’19, Association for Computing Machinery, New
  York, NY, USA (2019). \doi{10.1145/3292500.3330691}

\bibitem{search-neutrality}
Grimmelmann, J.: Some skepticism about search neutrality. The Next Digital
  Decade: Essays on the Future of the Internet  (2011)

\bibitem{disp-treatment}
Heidari, H., Krause, A.: Preventing disparate treatment in sequential decision
  making. In: IJCAI (2018)

\bibitem{chawla2019multicategory}
Ilvento, C., Jagadeesan, M., Chawla, S.: Multi-category fairness in sponsored
  search auctions. In: Proceedings of the 2020 Conference on Fairness,
  Accountability, and Transparency. FAT* ’20, Association for Computing
  Machinery, New York, NY, USA (2020). \doi{10.1145/3351095.3372848}

\bibitem{kuhlman_fare_2019}
Kuhlman, C., VanValkenburg, M., Rundensteiner, E.: Fare: Diagnostics for fair
  ranking using pairwise error metrics. In: The World Wide Web Conference. WWW
  ’19, Association for Computing Machinery, New York, NY, USA (2019).
  \doi{10.1145/3308558.3313443}

\bibitem{mehrotra_towards_2018}
Mehrotra, R., McInerney, J., Bouchard, H., Lalmas, M., Diaz, F.: Towards a fair
  marketplace: Counterfactual evaluation of the trade-off between relevance,
  fairness \& satisfaction in recommendation systems. In: Proceedings of the
  27th ACM International Conference on Information and Knowledge Management.
  CIKM ’18, Association for Computing Machinery, New York, NY, USA (2018).
  \doi{10.1145/3269206.3272027}

\bibitem{click-estimate}
Richardson, M., Dominowska, E., Ragno, R.: Predicting clicks: Estimating the
  click-through rate for new ads. In: Proceedings of the 16th International
  Conference on World Wide Web. WWW ’07, Association for Computing Machinery,
  New York, NY, USA (2007). \doi{10.1145/1242572.1242643}

\bibitem{sakai_simple_2010}
Sakai, T., Craswell, N., Song, R., Robertson, S.E., Dou, Z., Lin, C.Y.: Simple
  evaluation metrics for diversified search results. In: EVIA@NTCIR (2010)

\bibitem{serbos-fairness-2017}
Serbos, D., Qi, S., Mamoulis, N., Pitoura, E., Tsaparas, P.: Fairness in
  package-to-group recommendations. In: Proceedings of the 26th International
  Conference on World Wide Web. WWW ’17, International World Wide Web
  Conferences Steering Committee, Republic and Canton of Geneva, CHE (2017).
  \doi{10.1145/3038912.3052612}

\bibitem{Singh-Joachims}
Singh, A., Joachims, T.: Fairness of exposure in rankings. In: Proceedings of
  the 24th ACM SIGKDD International Conference on Knowledge Discovery \& Data
  Mining. KDD ’18, Association for Computing Machinery, New York, NY, USA
  (2018). \doi{10.1145/3219819.3220088}

\bibitem{fairness-classification}
Verma, S., Rubin, J.: Fairness definitions explained. In: Proceedings of the
  International Workshop on Software Fairness. FairWare ’18, Association for
  Computing Machinery, New York, NY, USA (2018). \doi{10.1145/3194770.3194776}

\bibitem{wan_addressing_2020}
Wan, M., Ni, J., Misra, R., McAuley, J.: Addressing marketing bias in product
  recommendations. In: Proceedings of the 13th International Conference on Web
  Search and Data Mining. WSDM ’20, Association for Computing Machinery, New
  York, NY, USA (2020). \doi{10.1145/3336191.3371855}

\bibitem{Wu-2018}
Wu, Y., Zhang, L., Wu, X.: On discrimination discovery and removal in ranked
  data using causal graph. In: Proceedings of the 24th ACM SIGKDD International
  Conference on Knowledge Discovery \& Data Mining. KDD ’18, Association for
  Computing Machinery, New York, NY, USA (2018). \doi{10.1145/3219819.3220087}

\bibitem{julia-paper}
Yang, K., Stoyanovich, J.: Measuring fairness in ranked outputs. In:
  Proceedings of the 29th International Conference on Scientific and
  Statistical Database Management. SSDBM ’17, Association for Computing
  Machinery, New York, NY, USA (2017). \doi{10.1145/3085504.3085526}

\bibitem{zehlike2017}
Zehlike, M., Bonchi, F., Castillo, C., Hajian, S., Megahed, M., Baeza-Yates,
  R.: Fa*ir: A fair top-k ranking algorithm. In: Proceedings of the 2017 ACM on
  Conference on Information and Knowledge Management. CIKM ’17, Association
  for Computing Machinery, New York, NY, USA (2017).
  \doi{10.1145/3132847.3132938}

\end{thebibliography}
\end{document}